\documentclass{llncs}
\usepackage{graphicx}
\usepackage{amsmath}
\usepackage{amsfonts}
\usepackage{color,soul}

\begin{document}
\title{Synthesis of Majority Expressions through Primitive Function Manipulation}
\titlerunning{Synthesis of Majority Expressions}

\author{Evandro C. Ferraz\inst{1}
\and Jeferson de Lima Muniz\inst{1}
\and Alexandre C. R. da Silva\inst{1}
\and Gerhard W. Dueck\inst{2}
}
\authorrunning{E. C. Ferraz et al.}

\institute{Department of Electrical Engineering, FEIS - Univ. Estadual Paulista, Ilha Solteira, SP, Brazil
\and
Faculty of Computer Science, University of New Brunswick, Fredericton, NB, Canada
}

\maketitle 

\begin{abstract}
 Due to technology advancements and circuits miniaturization, the study of logic systems that can be applied to nanotechnology has been progressing steadily. 
 Among the creation of nanoeletronic circuits reversible and majority logic stand out. 
 This paper proposes the $MPC$ (Majority Primitives Combination) algorithm, used for majority logic synthesis. 
 The algorithm receives a truth table as input and returns a majority function that covers the same set of minterms. 
 The formulation of a valid output function is made with the combination of previously optimized functions. As cost criteria the algorithm searches for a function with the least number of levels, followed by the least number of gates, inverters, and gate inputs. In this paper it's also presented a comparison between the $MPC$ and the exact\_mig, currently considered the best algorithm for majority synthesis. The exact\_mig encode the exact synthesis of majority functions using the number of levels and gates as cost criteria. The $MPC$ considers two additional cost criteria, the number of inverters and the number of gate inputs, with the goal to further improve exact\_mig results. Tests have shown that both algorithms return optimal solutions for all functions with $3$ input variables. For functions with $4$ inputs, the $MPC$ is able to further improve 42,987 (66\%) functions and achieves equal results for 7,198 (11\%). For functions with $5$ input variables, out of a sample of 1,000 randomly generated functions, the $MPC$ further improved 477 (48\%) functions and achieved equal results for 112 (11\%).

\keywords{Majority Logic \and Primitive Functions \and Logic Synthesis.}
\end{abstract}
\section{Introduction}

Majority logic allows the creation of nanoelectronic circuits for several different technologies, which justifies the search for majority based algorithms that generates optimized circuits.
Among the first works that deal with majority logic, are Lindaman \cite{lindaman1960theorem}, Cohn \cite{cohn1961axiomatic}, and  Akers \cite{akers1961truth}. 
Lindaman \cite{lindaman1960theorem} proposed the first theorem for applying majority logic in binary decision problems, introducing the majority operator to classical boolean algebra. 
The theorem, shown in equation \ref{lindaman}, proposes a boolean function equivalent to a majority operation. 

\begin{equation}
\label{lindaman}
M(A,B,C) = A\cdot B + A\cdot C + B\cdot C
\end{equation}

Subsequently, a set of axioms that defines the majority algebra independently of the classical boolean algebra was presented in \cite{cohn1961axiomatic}, creating the basis for current majority algebra axiomatization ($\Omega$).

Moreover, the authors in \cite{zhang2004method} presented a method that performs the mapping of all $3$-input boolean functions into a $3$-dimensional cube, generating $13$ possible patterns, where each pattern has a different formula to convert a classical boolean function into a majority equivalent. 

Similarly, the authors in \cite{wang2011minimal} presented a method that uses a $4$-dimensional cube to map $4$-input functions, generating a total of $143$ representation patterns. All $143$ patterns also have a specific formula to find their equivalent majority functions.

In majority algebra, simplification algorithms based on primitive functions are widely used. Primitive functions are functions with at most $1$ majority gate in their optimized form. An algorithm that maps each of the primitive functions and uses the obtained maps to generate more complex functions was proposed in \cite{walus2004circuit}. The mapping of functions is realized with Karnaugh Maps, a graphical method proposed by Maurice Karnaugh in $1953$, which aims to simplify a classic boolean function by mapping its truth table \cite{karnaugh1953map}.

In \cite{mishra2017heuristic} a similar algorithm was developed, the $B2M$ (Boolean to Majority). The $B2M$ receives a boolean function as input and generates a majority function that covers the same set of minterms. The generation of an output function is also done with the combination of primitives, selected by their $MLD$ (Modified Levenshtein Distance). 

The authors in \cite{wang2015synthesis} proposed a methodology that combines lower level majority functions, starting from primitives, to form higher level majority functions. The goal of this method is to build a majority expressions lookup table ($MLUT$) that stores the majority equivalent for all possible $4$-input boolean functions. Using the $MLUT$, the algorithm will then search the equivalent majority expression for every node in the input network, generating a majority network as output.  

The authors in \cite{soeken2017exact} proposed the $exact\_mig$ algorithm, which is considered state of the art.
As input, the algorithm receives a truth table or a $MIG$ (Majority Inverter Graph) \cite{amaru2014majority}, with a maximum of $6$ input variables, and returns a majority function that covers the same set of minterms. The most important characteristic of this algorithm is the proposal of a exact synthesis for majority functions. The function is built from a set of constraints ($K$) that shape a given problem accordingly to the definitions of the majority boolean algebra. The majority output function is generated with the application of $K$ to an $SMT$ (Satisfiability Modulo Theory) solver \cite{de2008z3}. As cost criteria the $exact\_mig$ takes into consideration the number of levels and gates in the output function, making it possible to choose which of these criteria will be prioritized.

In this work the $MPC$ algorithm is proposed. Similiar to the methodology proposed in \cite{wang2015synthesis}, the algorithm checks all possible combinations among primitive functions and creates a table to store them. For each function, the covered set of minterms is also stored. If there are two functions that covers the same set of minterms, the lowest cost function is kept and the other function is discarded. As a result, we have a table ($M_2$) that lists all the sets covered by majority functions with $2$ levels. As cost criteria the algorithm considers the depth of the function, followed by the number of gates, the number of inverters, and the number of gate inputs in the output function.

The $MPC$ can be used to synthesize boolean functions with a maximum of $5$-input variables. For $3$-input variables the algorithm returns an optimal solution for all possible functions. For $4$ and $5$-input variables the algorithm guarantees an optimal solution for functions covered by $M_2$ or by a primitive, and uses a specific synthesis to cover functions with a higher number of levels. For $5$ variables however, functions with $4$ or more levels are generated by the application of the Shannon Theorem.  

This article is organized as follows: In section \ref{majority}, we present an explanation about majority algebra, including its axiomatization and the concept of primitive majority functions. Section  \ref{algorithm} presents the $MPC$ algorithm, explaining how it works for $3$, $4$ and $5$-input variables. Section \ref{results} presents the results obtained comparing the $MPC$ and the $exact\_mig$. Section \ref{conclusion} presents the conclusion of what was realized in the paper.

\section{Majority Boolean Algebra} \label{majority}

The majority boolean algebra is composed by the set $\left\{\mathbb{B},\neg,M\right\}$. The elements $\mathbb{B}$ and $\neg$, 
as in classical boolean algebra, represent the binary values $\left\{0,1\right\}$ and the inversion operator, respectively, and $M$ represents the majority operator \cite{chattopadhyay2016notes}.

A majority function returns as output the most present binary value among its inputs. Therefore, an operator $M$ that has a total of $3$-input variables will return a true value only if two or more inputs are true. The truth table presented in Table \ref{tab_TTMAJ} exemplifies a majority operation for the variables $X$, $Y$ and $Z$.

\begin{table}[ht]
\centering
\caption{Example of a majority operation.}
\label{tab_TTMAJ}
\begin{tabular}{|c|c|c|c|}
\hline
\textbf{$X$} & \textbf{$Y$} & \textbf{$Z$} & \textbf{$M(X,Y,Z)$} \\ \hline
0 & 0 & 0 & 0 \\ \hline
0 & 0 & 1 & 0 \\ \hline
0 & 1 & 0 & 0 \\ \hline
0 & 1 & 1 & 1 \\ \hline
1 & 0 & 0 & 0 \\ \hline
1 & 0 & 1 & 1 \\ \hline
1 & 1 & 0 & 1 \\ \hline
1 & 1 & 1 & 1 \\ \hline
\end{tabular}
\end{table}  

From a majority operation it's also possible to obtain $AND$ and $OR$ functions, performed by fixing one of the input variables to a constant binary value.

As an example, the function $M(A,B,C)$ is considered. Setting the value of $A$ to $0$, we have an $AND$ function between $B$ and $C$. Setting the value of $A$ to $1$, we have an $OR$ function between $B$ and $C$. This example is shown in the Table \ref{tab_TTMAND}.

\begin{table}[ht]
\centering
\caption{Generation of functions $AND$ and $OR$.}
\label{tab_TTMAND}
\begin{tabular}{|c|c|c|c|c|c|}
\hline
\textbf{$B$} & \textbf{$C$} & \textbf{$B \cdot C$} & \textbf{$M(0,B,C)$} & \textbf{$B + C$} & \textbf{$M(1,B,C)$} \\ \hline
0 & 0 & 0 & 0 & 0 & 0 \\ \hline
0 & 1 & 0 & 0 & 1 & 1 \\ \hline
1 & 0 & 0 & 0 & 1 & 1 \\ \hline
1 & 1 & 1 & 1 & 1 & 1 \\ \hline
\end{tabular}
\end{table}  

\subsection{Axiomatization of majority functions ($\Omega$)}

The set of axioms that defines the majority algebra is represented by $\Omega$ and can be divided into axioms of Commutativity, Associativity, Distribution, Inverter Propagation and Majority \cite{amaru2016sound}.

The Commutativity axiom ($\Omega.C$), represented in Equation \ref{OC}, determines that the input order doesn't change the output value.

\begin{equation}
\label{OC}
M(X,Y,Z) = M(X,Z,Y) =  M(Z,Y,X)
\end{equation}

The Associativity axiom ($\Omega$.A) states that the exchange of variables between two functions is possible, as long as they are at subsequent levels and have one variable in common. An example of an $\Omega$.A application is presented in Equation \ref{AC}.

\begin{equation}
\label{AC}
M(X,U,M(Y,U,Z)) = M(Z,U,M(Y,U,X)) 
\end{equation}

Note that the variable shared between levels is $U$. Therefore, it's possible to substitute the remaining variable in the upper level for one in the subsequent level. In the presented example, we had an exchange between the variables $X$ and $Z$.

The Distribution axiom ($\Omega$.D) determines that it's possible to distribute a set of variables to gates in subsequent levels. In Equation \ref{OD} an example of this theorem is given, where the distributed set is $\left\{X,Y\right\}$.

\begin{equation}
\label{OD}
\scalebox{0.85}{
M(X,Y,M(U,V,Z)) = M(M(X,Y,U), M(X,Y,V), Z)}
\end{equation}

The Inverter propagation axiom ($\Omega$.I), represented in Equation \ref{OI1}, determines that a majority function is self-dual \cite{akers1962synthesis}. 

\begin{equation}
\label{OI1}
\overline{M}(X,Y,Z) = M(\overline{X},\overline{Y},\overline{Z})
\end{equation}

The Majority ($\Omega$.M) can be divided in $2$ equations. Equation \ref{OM1} shows that the output of a majority gate is equal to the most common value among its inputs. Equation \ref{OM2} shows that the output value will be equal to the tie-breaking variable in functions with the same number of true and false values.

\begin{equation}
\label{OM1}
M(X,X,Y) = X
\end{equation}

\begin{equation}
\label{OM2}
M(X,\overline{X},Y) = Y
\end{equation}

\subsection{Primitive Majority Functions}
 
Primitive functions can be obtained by a single gate. In the majority algebra, primitive functions (also called primitives) can be used as a base for the construction of more complex functions. All primitives can be obtained from the sets $C$, $V$, $G$ and $T$, where each set corresponds to functions with a specific number of inputs. The total number of primitives is obtained by summing the functions in $C$, $V$, $G$ and $T$ \cite{wang2015synthesis}.

The set $C$ represents functions with no input variables, covering the constants $0$ and $1$. Therefore, $|C|= 2$.

The set $V$ represents all functions formed by a single input variable, in its complemented form or not. Equation \ref{VC} shows how to calculate the number of functions in $V$.

\begin{equation}
\label{VC}
|V| = 2 \cdot n 
\end{equation}

In Table \ref{tab_TTV}, we can observe the listing of $V$ for $3$ input variables. The number of input variables are represented by $n$. Note that the classical functions and their corresponding majority forms are equal because the $V$ set is composed only by functions without operators. 

\begin{table}[ht]
\centering
\caption{List of set $V$ for $n = 3$.}
\label{tab_TTV}
{\renewcommand{\arraystretch}{1.2}
\begin{tabular}{|c|c|}
\hline
\textbf{Classic Function} & \textbf{Majority Function} \\ \hline
$A$ & $A$ \\ \hline
$B$ & $B$ \\ \hline
$C$ & $C$ \\ \hline
$\overline{A}$ & $\overline{A}$ \\ \hline
$\overline{B}$ & $\overline{B}$ \\ \hline
$\overline{C}$ & $\overline{C}$ \\ \hline
\end{tabular}
}
\end{table} 

The set $G$ is formed by functions with a single $AND$ or $OR$ operator, having a total of $2$ input variables. The number of functions in $G$ can be calculated by the Equation \ref{G}. The variables $E$ and $O$ represents the possible combinations of inputs, for $AND$ and $OR$ operations respectively. For $n = 3$, we have $E = \left\{A\cdot B,A\cdot C,B\cdot C\right\}$ and $O = \left\{A+B,A+C,B+C\right\}$. Each combination has $4$ inversion variations, the combination $A + B$ for example, has the variations $\left\{A+B,\overline{A}+B,A+\overline{B},\overline{A}+\overline{B}\right\}$.

\begin{equation}
\label{G}
|G| = (4 \cdot |E|) + (4 \cdot |O|) 
\end{equation} 

In Table \ref{tab_TTG}, we present the functions in $G$ for $n = 3$. 

\begin{table}[ht]
\centering
\caption{List of set $G$ for $n = 3$.}
\label{tab_TTG}
{\renewcommand{\arraystretch}{1.2}
\begin{tabular}{|c|c|}
\hline
\textbf{Classic Function} & \textbf{Majority Function} \\ \hline
$A\cdot B$ & $M(A,B,0)$ \\ \hline
$\overline{A}\cdot B$ & $M(\overline{A},B,0)$ \\ \hline
$A\cdot \overline{B}$ & $M(A,\overline{B},0)$ \\ \hline
$\overline{A} \cdot \overline{B}$ & $\overline{M}(A,B,1)$ \\ \hline
$A\cdot C$ & $M(A,0,C)$ \\ \hline
$\overline{A}\cdot C$ & $M(\overline{A},0,C)$ \\ \hline
$A\cdot \overline{C}$ & $M(A,0,\overline{C}$) \\ \hline
$\overline{A}\cdot \overline{C}$ & $\overline{M}(A,1,C)$ \\ \hline
$B\cdot C$ & $M(0,B,C)$ \\ \hline
$\overline{B}\cdot C$ & $M(0,\overline{B},C)$ \\ \hline
$B\cdot \overline{C}$ & $M(0,B,\overline{C}$) \\ \hline
$\overline{B}\cdot \overline{C}$ & $\overline{M}(1,B,C)$ \\ \hline
$A+B$ & $M(A,B,1)$ \\ \hline
$\overline{A}+B$ & $M(\overline{A},B,1)$ \\ \hline
$A+\overline{B}$ & $M(A,\overline{B},1)$ \\ \hline
$\overline{A}+\overline{B}$ & $\overline{M}(A,B,0)$ \\ \hline
$A+C$ & $M(A,1,C)$ \\ \hline
$\overline{A}+C$ & $M(\overline{A},1,C)$ \\ \hline
$A+\overline{C}$ & $M(A,1,\overline{C}$) \\ \hline
$\overline{A}+\overline{C}$ & $\overline{M}(A,0,C)$ \\ \hline
$B+C$ & $M(1,B,C)$ \\ \hline
$\overline{B}+C$ & $M(1,\overline{B},C)$ \\ \hline
$B+\overline{C}$ & $M(1,B,\overline{C}$) \\ \hline
$\overline{B}+\overline{C}$ & $\overline{M}(0,B,C)$ \\ \hline
\end{tabular}
}
\end{table} 

The set $T$ represents functions with a single majority gate, with no constant value and no repeated variable as input. Equation \ref{T} calculates the number of functions in $T$. The variable $t$ represents the number of possible combinations among the input variables, considering $3$ inputs per combination. Note that each combination has $8$ variations of inverters and, for $n = 3$, there is only one possible combination.
\begin{equation}
\label{T}
|T| = t \cdot 8 
\end{equation} 

Table \ref{tab_TT} shows the list of functions in $T$, for $n = 3$.

\begin{table}[ht]
\centering
\caption{List of functions in $T$ for $n = 3$.}
\label{tab_TT}
{\renewcommand{\arraystretch}{1.2}
\begin{tabular}{|c|c|}
\hline
\textbf{Classic Function} & \textbf{Majority Function} \\ \hline
$AB + AC + BC$ & $M(A,B,C)$ \\ \hline
$\overline{A}\cdot \overline{B} + \overline{A}\cdot \overline{C}+\overline{B}\cdot \overline{C}$ & $\overline{M}(A,B,C)$ \\ \hline
$\overline{A}\cdot B + \overline{A}\cdot C + B\cdot C$ & $M(\overline{A},B,C)$ \\ \hline
$A\cdot \overline{B} + A\cdot \overline{C}+\overline{B}\cdot \overline{C}$ & $\overline{M}(\overline{A},B,C)$ \\ \hline
$A\cdot \overline{B} + A\cdot C +\overline{B}\cdot C$ & $M(A,\overline{B},C)$ \\ \hline
$\overline{A}\cdot B + \overline{A}\cdot \overline{C} + B\cdot \overline{C}$ & $\overline{M}(A,\overline{B},C)$ \\ \hline
$A\cdot B + A\cdot \overline{C} + B\cdot \overline{C}$ & $M(A,B,\overline{C}$) \\ \hline
$\overline{A}\cdot \overline{B} + \overline{A}\cdot C +\overline{B}\cdot C$& $\overline{M}(A,B,\overline{C})$ \\ \hline
\end{tabular}
}
\end{table} 

\section{The $MPC$ Algorithm}  \label{algorithm}

In this section we propose the $MPC$ algorithm. The $MPC$ receives a truth table $f$ as input and returns a majority function that covers the same set of minterms. To generate a valid output function we use the expression $M(X_1,X_2,X_3)$.
Each variable $X_c$, where $1 \leq c \leq 3$, represents a majority primitive or a $2$-level majority function.

\subsection{Tables Formulation}

The first step of $MPC$ is the tables formulation phase, where the functions used to build $M(X_1,X_2,X_3)$ are formulated. The algorithm receives an input truth table $f$, indentifies the number of input variables, represented by $n$, and generates the primitives table based on the sets $C$, $V$, $G$ and $T$. We also store the set of minterms covered by every primitive function. Note that each primitive function is the optimal solution of its respective set of minterms.

The second table build by the $MPC$ is the $M_2$ table, formed by the aplication of all possible combinations among primitive functions in the expression $M(X_1,X_2,X_3)$, without considering repeated primitives. For each generated function the set of covered minterms is also stored. If a set is covered by $2$ or more functions, the one wich the lowest cost is kept and the others are discarded. Therefore, the table $M_2$ lists all sets of minterms that can be covered by a $2$-level majority function and, since they are obtained exhaustively, $M_2$ functions are also an optimal solution for their respective set of minterms. It is also important to point that, for computational performance optimization, the $M_2$ is stored as a $LUT$ (Look-Up Table) in the $MPC$ code.  

As an example of a $M_2$ function, we have $M(X_1,X_2,X_3) = M(A,M(A,\overline{B},0),\overline{M}(A,B,C))$, where $X_1 = A$, $X_2 = M(A,\overline{B},0)$ and $X_3 = \overline{M}(A,B,C)$.

The cost criteria used by $MPC$ is primarily the number of levels and gates in the output function, followed by the number of inverters and gate inputs. 

To ensure the minimization of inverters, the $1$ gate primitives follow $4$ possible paterns: 

\begin{itemize}
  \item $M(A,B,C)$, no inverters; 
  \item $M(\overline{A},B,C)$, a single complemented input; 
  \item $\overline{M}(A,B,C)$, only $1$ inverter applied to the output value;
  \item $\overline{M}(\overline{A},B,C)$, $1$ input and the output complemented.
\end{itemize}

Note that in cases where the gate has $2$ inverters, even thought the number of inverters stay the same, it's better to negate the output and only $1$ input, since $M(\overline{X},Y,\overline{Z}) = \overline{M}(X,\overline{Y},Z)$. This allows the application of $\Omega.I$ to minimize the number of inverters when the primitives are being used to build functions with more than $1$ level. To exemplify this application we consider: $M(\overline{M}(A,\overline{B},C),\overline{D},0)$, wich has $2$ levels, $2$ gates and $3$ inverters. By applying $\Omega.I$ we have $M(\overline{M}(A,\overline{B},C),\overline{D},0) = \overline{M}(M(A,\overline{B},C),D,1)$, wich has the same number of levels and gates, but has $1$ less inverter.  

The total of possible functions for a specific number of inputs is represented by the variable $S$, and can be calculated by $2^{m}$. Note that $m = 2^{n}$, and represents the number of terms in the input truth table $f$.

For $n=3$, $S = 256$. The primitives table covers $40$ of these functions. The $216$ left are covered by the $M_2$ table. Therefore, $S$ can be completely covered by majority expressions with at most $2$ levels, which makes the table formulation phase enough for obtaining all optimal solutions for $n=3$.

For $n=4$, $S$ = 65,536 and $90$ of these functions are primitives, with at most $1$ majority gate. In the formulation of $M_2$, only 10,260 functions can be covered. For the remaining 55,186, 55,184 can be covered by majority expressions with $3$ levels. The remaining $2$ functions need a majority expression with $4$ levels to be covered. 

\subsection{$MPC$ Synthesis for $4$-input functions}

This section presents the synthesis used in $MPC$ for the construction of majority functions where $n=4$. The objective of this synthesis is to formulate $M(X_1,X_2,X_3)$ with the combination of primitives and $M_2$ functions, generating a majority function that covers the same minterms of $f$. Note that this synthesis is only applied if $f$ can't be covered by any function in the $M_2$ table or by any primitive.

The synthesis is composed by two different loops, each one having their own characteristics. If a output function couldn't be found in the first loop the second starts.

The first loop is composed by the following steps:

\begin{enumerate}

\item Any primitive or $M_2$ function that doesn't cover at least one minterm of $f$ is discarded from its respective table;

\item Build a new table $P$, selecting every pair of primitives $(p_1 + p_2)$ where:

\begin{itemize}
  \item Every minterm in $f$ is covered at least once by $p_1 + p_2$; 
  \item The pair $p_1 + p_2$ only covers minterms of $f$.
\end{itemize}

\item Select a pair of primitives from $P$, as $X_1$ and $X_2$;

\item Create a vector $v$ with $2^{n}$ elements, that will be used to build the  truth table for $X_3$. 
 Every element in $v$ represents a minterm in $f$.
The vector $v$ is updated according to the set of minterms covered by $X_1$ and $X_2$. If a minterm $i$ is covered by both functions, $v_{i} = 2$. If it's covered by only one function, $v_{i} = 1$. And if it isn't covered by any function, $v_{i} = 0$.
For  example, given  $f = \{0,1,5,8\}$, $X_1 = \{0,1,4,5\}$ and $X_2 = \{0,1,2,8,10\}$. Then $v$ has the values as shown in Table \ref{tab_v}.

\begin{table}[ht]
\centering
\caption{Generation of vector $v$.}
\label{tab_v}
\begin{tabular}{|c|c|c|c|c|}
\hline
\textbf{ Minterms } & \textbf{$f = \{0,1,5,8\}$} & \textbf{$X_1 = \{0,1,4,5\}$} & \textbf{$X_2 = \{0,1,2,8,10\}$} & \textbf{  $v$  } \\ \hline
0 & 1 & 1 & 1 & 2 \\ \hline
1 & 1 & 1 & 1 & 2 \\ \hline
2 & 0 & 0 & 1 & 1 \\ \hline
3 & 0 & 0 & 0 & 0 \\ \hline
4 & 0 & 1 & 0 & 1 \\ \hline
5 & 1 & 1 & 0 & 1 \\ \hline
6 & 0 & 0 & 0 & 0 \\ \hline
7 & 0 & 0 & 0 & 0 \\ \hline
8 & 1 & 0 & 1 & 1 \\ \hline
9 & 0 & 0 & 0 & 0 \\ \hline
10 & 0 & 0 & 1 & 1 \\ \hline
11 & 0 & 0 & 0 & 0 \\ \hline
12 & 0 & 0 & 0 & 0 \\ \hline
13 & 0 & 0 & 0 & 0 \\ \hline
14 & 0 & 0 & 0 & 0 \\ \hline
15 & 0 & 0 & 0 & 0 \\ \hline
\end{tabular}
\end{table} 

\item Create the  truth table for $X_3$, represented by the vector $X_3f$.
Positions where $v_{i} = 2$ or $v_{i} = 0$ are considered as don't care states (represented by $x$).
For positions where $v_{i} = 1$ and $i$ is also covered by $f$, we have $X_3f_i = 1$. If $v_{i} = 1$ and $i$ isn't covered by $f$, we have $X_3f_i = 0$. Therefore, for the example presented in Table \ref{tab_v}, we have $X_3f = [xx0x01xx1x0xxxxx]$;    

\item Generate every possible truth table manipulating the don't care states in $X_3f$. Each possibility is searched in the $M2$ table. From the functions, a new table, $P_3$, is constructed.

\item For every function in $P_3$ composed by a gate that also composes $X_1$ or $X_2$, we reduce its cost by $1$. This rule exists because each gate is counted only once in the calculation of a majority function size.

\item Select the lowest cost function in $P_3$, that hasn't been selected yet, as $X_3$. If there's no valid $X_3$, we go back to step $3$ and find a new primitive pair.

\item With the selection of $X_3$ we now have a valid output $M(X_1,X_2,X_3)$. To minimize inverters,  $\Omega.I$ is applied in every level of the function built. If the function post $\Omega.I$ application has a lower cost, the previous function is substituted.

\item The loop terminates when every possible primitive pairs in $P$ with a function from $M_2$, and every $M(X_1,X_2,X_3)$ found is stored in a table $Z$, have been combined.

\item By the end of the loop, the algorithm returns the function with the lowest cost in $Z$. If no function could be found the second loop starts.

\end{enumerate}

From all $55,184$ sets of minterms that can be covered by a $3$ level function, $50,016$ can be covered by functions where $2$ elements of $X_c$ are primitives. Those functions are found by the first loop. 

Among the $5,168$ remaining sets, $5,056$ can be covered by functions where only $1$ element of $X_c$ is a primitive. The $112$ remaining sets that can only be covered by functions where all elements of $X_c$ are $2$ level functions from $M_2$. Those functions are found by the second loop.

The second loop is composed by the following steps:

\begin{enumerate}

\item Select $X_1$ from the primitives table. If every primitive function has been selected as $X_1$ and a valid output function could not be found, $X_1$ is selected from a group of functions $R$. The group $R$ is formed by every $M_2$ function with size $r$, where $r$ represents the number of gates in a $M_2$ function. Therefore, $r$ starts at $2$, the lowest number of gates that a $2$ level majority function can have, and is incremented if a group $R$ with higher size functions must be defined.  

\item Create $2$ new vectors, $v_{0}$ and $v_{-1}$. The vector $v_{0}$ contains the positions of $f$ that haven't been covered yet, therefore $v_{0} = f - X_1$. The vector $v_{-1}$ has the positions of $v$ that can't be covered one more time, therefore $v_{-1} = X_1 - f$;

\item From $v_{0}$ and $v_{-1}$  the  truth tables for $X_2$, represented by the variable $X_2f$ are generated.
 $X_2f$ represents a truth table, with the same size of $f$, that can have binary values or don't care states.
For the minterms stored in $v_{0}$, $X_2f_i = 1$. For minterms stored in $v_{-1}$, $X_2fi = 0$. The other minterms are all considered don't care states.

\item Every possible truth table manipulating the don't care states in $X_2f$ is generated. 
Each possibility is searched in the $M_2$ table. From these functions a new table, $P_2$, is created.

\item For every function in $P_2$ that is composed by a gate that also composes $X_1$, its cost is reduced by $1$.

\item Select the lowest cost function in $P_2$, that was not selected yet, as $X_2$. If there's no valid $X_2$,  go back to step $1$ and select a new $X_1$.

\item To find $X_3$ create $X_3f$ based on $v_{-1}$ and a new vector $v_{1}$. 
The vector $v_{1}$ stores the minterms of $f$ covered only once by $X_c$. Therefore, the minterms in $v_{1}$ must be covered by $X_3$. For the minterms stored in $v_{-1}$, $X_3f = 0$. For the minterms stored in $v_{1}$, $X_3f = 1$.

\item To find all possibilities for $X_3f$, search the respective functions in the $M_2$ table and  build $P_3$ from them.

\item Again,  update the cost of the functions in $P_3$ based on the gates in $X_1$ and $X_2$.

\item Select the lowest cost function in $P_3$, that hasn't been selected yet, as $X_3$. 
If there's no valid $X_3$,  go back to the step $6$ and select a new $X_2$.

\item With the selection of $X_3$ we now have a valid output $M(X_1,X_2,X_3)$. For the minimization of inverters we also apply $\Omega.I$ in every level of the function built and we substitute it if the function post $\Omega.I$ application has a lower cost.

\item Every $M(X_1,X_2,X_3)$ found is stored in the table $Z$ and the loop stops when all primitive functions are selected as $X_1$. If no function could be found, the algorithm goes back to step $1$ and restarts selecting $X_1$ from a group $R$, stoping when all functions in $R$ were selected as $X_1$. If yet no function could be found, the algorithm increments $r$ and restarts the loop with a new group $R$. The algorithm returns the lowest cost function stored in $Z$ as output. 

\end{enumerate}

For the $2$ sets that need a function with $4$ levels to be covered, we first select $X_1$ from the primitives table, then we build $X_2$ and $X_3$ as $3$ level functions using the explained synthesis.

\subsection{$MPC$ Synthesis for $5$-input functions}

The synthesis for $5$-input ($n=5$) functions also uses the primitives and the $M_2$ table as a base to build functions with a higher number of levels. 

For $n=5$, $S = 4,294,967,296$ and $172$ of these sets can be covered by primitives, with at most $1$ majority gate. The $M_2$ table stores the $253,560$ sets that can be covered by majority functions with $2$ levels. The remaining sets needs more than $2$ levels to be covered.

To build $3$ level functions the algorithm also uses the expression $M(X_1,X_2,X_3)$, realizing the combination of primitives and $M_2$ functions, selected by their lowest cost.

The complete synthesis for $3$ level functions is composed by the following steps:

\begin{enumerate} 

\item Order by cost every function from the primitives and $M_2$ tables.

\item Select the function with the lowest cost as $X_1$.

\item Reduce the cost by 1 for every primitive or $M_2$ function that is composed by a gate that also composes $X_1$.

\item Create $v_{0}$ and $v_{-1}$, where $v_{0} = f - X_1$ and $v_{-1} = X_1 - f$.

\item We select $X_2$, primarly from the primitives, as the lowest cost function that:

\begin{itemize} 
\item Covers all minterms in $v_{0}$.

\item Doesn't cover any minterm of $v_{-1}$.
\end{itemize}

If no valid $X_2$ can be found among the primitives, select $X_2$ from the $M_2$ table. If still no valid $X_2$ can be found,  go back to the step $2$ and select a new $X_1$.

\item Again, update the cost of the primitives and $M_2$ functions based on the gates in $X_1$ and $X_2$.

\item Create $v_{1}$, where $v_{1}$ stores the minterms of $f$ covered only once by $X_c$.

\item Select $X_3$, from the primitives, the lowest cost function that:

\begin{itemize} 
\item Covers all minterms in $v_{1}$.

\item Doesn't cover any minterm of $v_{-1}$.
\end{itemize}

If no valid $X_3$ could be found among the primitives, we select $X_3$ from the $M_2$ table. If still no valid $X_3$ could be found, we go back to the step $5$ and select a new $X_2$.

\item With the selection of $X_3$ we now have a valid output. Next apply $\Omega.I$ in every level of $M(X_1,X_2,X_3)$ and return the lowest cost version as output.

\end{enumerate}

For functions that needs more than $3$ levels to be covered we apply the reduction of fan-ins by Shannon expansion. 
Equation $\ref{shanon}$ shows the equivalent majority version of the Shannon theorem, applied to the set of inputs $\{A,B,C,D,E\}$.

\begin{equation}
\label{shanon}
M(A,B,C,D,E) = M(M(F_1,0,A),M(F_2,0,\overline{A}),1)
\end{equation}

The variable $A$ represents the isolated variable and $F_1$ and $F_2$ represent functions built with the remaining inputs \{B,C,D,E\}. 

The first step to apply this equation in the $MPC$ algorithm is to isolate the first input ($A$).
Then split the input truth table $f$ in two pieces to form two new truth tables, $f_1$ and $f_2$.  

Table \ref{shannonTT} shows an example of $f_1$'s and $f_2$'s generation. For this example, 
$f$ = [01110001100010100000111001010101] and the set of inputs are $\{A,B,C,D,E\}$ ($n=5$).

Note that, by splitting $f$ in two equal size tables, we have $f_1$ = [0111000110001010] and $f_2$ = [0000111001010101], where the set of inputs becames $\{B,C,D,E\}$ ($n=4$) and the variable $A$ is isolated.

\begin{table}[ht]
\center
\caption{Example of $f_1$'s and $f_2$'s generation by Shannon Theorem.}
\label{shannonTT}
\begin{tabular}{|c|c|c|c|c|c|c|}
\hline 
\textbf{Minterms} & \textbf{ B } & \textbf{ C } & \textbf{ D } & \textbf{ E } & \textbf{ $f_1$ } & \textbf{ $f_2$ } \\ \hline
\textbf{0} & $0$ & $0$ & $0$ & $0$ & $0$ & $0$ \\ \hline
\textbf{1} & $0$ & $0$ & $0$ & $1$ & $1$ & $0$ \\ \hline
\textbf{2} & $0$ & $0$ & $1$ & $0$ & $1$ & $0$ \\ \hline
\textbf{3} & $0$ & $0$ & $1$ & $1$ & $1$ & $0$ \\ \hline
\textbf{4} & $0$ & $1$ & $0$ & $0$ & $0$ & $1$ \\ \hline
\textbf{5} & $0$ & $1$ & $0$ & $1$ & $0$ & $1$ \\ \hline
\textbf{6} & $0$ & $1$ & $1$ & $0$ & $0$ & $1$ \\ \hline
\textbf{7} & $0$ & $1$ & $1$ & $1$ & $1$ & $0$ \\ \hline
\textbf{8} & $1$ & $0$ & $0$ & $0$ & $1$ & $0$ \\ \hline
\textbf{9} & $1$ & $0$ & $0$ & $1$ & $0$ & $1$ \\ \hline
\textbf{10} & $1$ & $0$ & $1$ & $0$ & $0$ & $0$ \\ \hline
\textbf{11} & $1$ & $0$ & $1$ & $1$ & $0$ & $1$ \\ \hline
\textbf{12} & $1$ & $1$ & $0$ & $0$ & $1$ & $0$ \\ \hline
\textbf{13} & $1$ & $1$ & $0$ & $1$ & $0$ & $1$ \\ \hline
\textbf{14} & $1$ & $1$ & $1$ & $0$ & $1$ & $0$ \\ \hline
\textbf{15} & $1$ & $1$ & $1$ & $1$ & $0$ & $1$ \\ \hline
\end{tabular}
\end{table}

To find $F_1$ and $F_2$ we apply the $MPC$ synthesis for $n=4$, explained in the previous section, to $f_1$ and $f_2$ respectively.

Note that the functions built by the Shannon Theorem aren't an optimal solution for $f$, since Equation \ref{shanon} adds $2$ levels and $3$ gates by itself.

\section{Results} \label{results}

In this section results obtained from the comparison of the algorithms $MPC$ and \textit{exact\_mig} are presented. 
For $n=4$ both algorithms were executed for all 65,536 possible functions. The obtained results were then compared based on the cost criteria used by the $MPC$, that prioritizes first the number of levels in the output function, followed by the number of gates, the number of inverters, and the number of gate inputs. In Table \ref{Compara} each column corresponds to a group $S_i$, where $0 \leq i \leq 2^{n}$.

Each $S_i$ represents a total of functions with a specific number of minterms. $S_4$, for example, represents every function that covers $4$ minterms among the 65,536 possibilities. For each $S_i$ the table shows the quantity of functions where $MPC$ generated results with a lower, higher and equal cost than \textit{exact\_mig}.

\begin{table}[ht]
\raggedleft
\caption{Cost comparison between $MPC$ and \textit{exact\_mig}.}
\label{Compara}
{\renewcommand{\arraystretch}{1.2}
\begin{tabular}{|c|c|c|c|c|}
\hline 
\textbf{\textit{i}} & \textbf{$S_{i}$} & \textbf{$MPC <$ \textit{exact\_mig}} & \textbf{$MPC >$ \textit{exact\_mig}} & \textbf{$MPC$ = \textit{exact\_mig}}  \\ \hline
\textbf{0} & 1 & 0 & 0 & 1 \\ \hline
\textbf{1} & 16 & 16 & 0 & 0 \\ \hline
\textbf{2} & 120 & 41 & 8 & 71 \\ \hline
\textbf{3} & 560 & 324 & 60 & 176 \\ \hline
\textbf{4} & 1,820 & 808 & 708 & 304 \\ \hline
\textbf{5} & 4,368 & 2,906 & 583 & 879 \\ \hline
\textbf{6} & 8,008 & 4,493 & 2,276 & 1,239 \\ \hline
\textbf{7} & 11,440 & 7,188 & 3,300 & 952 \\ \hline
\textbf{8} & 12,870 & 8,108 & 3,474 & 1,288 \\ \hline
\textbf{9} & 11,440 & 7,536 & 3,022 & 882 \\ \hline
\textbf{10} & 8,008 & 6,273 & 1,121 & 614 \\ \hline
\textbf{11} & 4,368 & 3,334 & 512 & 522 \\ \hline
\textbf{12} & 1,820 & 1,373 & 279 & 168 \\ \hline
\textbf{13} & 560 & 482 & 0 & 78 \\ \hline
\textbf{14} & 120 & 93 & 8 & 19 \\ \hline
\textbf{15} & 16 & 12 & 0 & 4 \\ \hline
\textbf{16} & 1 & 0 & 0 & 1 \\ \hline
\textbf{TOTAL} & \textbf{65,536} & \textbf{42,987} & \textbf{15,351} & \textbf{7,198} \\ \hline
\end{tabular}
}
\end{table} 

The $MPC$ generates lower cost results for $42,987 (66\%)$ functions, generates results with equal cost for $7,198 (11\%)$ functions and generates results with higher cost for the remaining $15,351 (23\%)$.
Note that $MPC$ is able to generate better results because \textit{exact\_mig} aims for the exact synthesis of only depth and size, while $MPC$ considers also the number of inverters and the number of gate inputs as cost criteria. In this comparison, the \textit{exact\_mig} functions where generated with the prioritization of depth, followed by the function size, differing from $MPC$ only by the addition of the number of inverters and gate inputs as third and forth criteria, respectively.

In Tables \ref{CompRunTimeSi} and \ref{CompRunTimeLvl}, comparisons about the runtime of both algorithms are presented. Table \ref{CompRunTimeSi} shows the total and average runtime for every $S_{i}$. Table \ref{CompRunTimeLvl} shows the total and average runtime for all functions with a specific depth. The comparisons were made in a computer with 8GB RAM and a 1.7GHZ CPU.

\begin{table}[ht]
\raggedleft
\caption{Runtime comparison between $MPC$ and \textit{exact\_mig} by $S_{i}$}
\label{CompRunTimeSi}
\begin{tabular}{|c|c|c|c|c|c|}
\hline
\multicolumn{2}{|c|}{} & \multicolumn{2}{c|}{\textbf{MPC}}      & \multicolumn{2}{c|}{\textbf{exact\_mig}} \\ \hline
\textit{i} & $S_{i}$   & Total Runtime     & Avg. Runtime       & Total Runtime 	& Avg. Runtime  \\ \hline
0          & 1         &  0,01 sec         &  0,01 sec          &  0,01 sec     	&  0,01 sec      \\ \hline
1          & 16        &  1,10 sec         &  0,06 sec    		&  1,65 sec        	&  0,10 sec       \\ \hline
2          & 120       &  30,02 sec        &  0,25 sec          &  17,85 sec    	&  0,14 sec       \\ \hline
3          & 560       &  8,29 min         &  0,88 sec       	&  4,57 min      	&  0,49 sec       \\ \hline
4          & 1,820     &  23,41 min    	   &  0,77 sec          &  21,44 min     	&  0,70 sec       \\ \hline
5          & 4,368     &  3,10 h       	   &  2,55 sec       	&  2,09 h        	&  1,72 sec       \\ \hline
6          & 8,008     &  11,39 h      	   &  5,12 sec       	&  3,84 h        	&  1,73 sec       \\ \hline
7          & 11,440    &  24,76 h          &  7,79 sec          &  21,74 h          &  6,84 sec       \\ \hline
8          & 12,870    &  19,64 h          &  5,49 sec          &  9,36 h           &  2,61 sec       \\ \hline
9          & 11,440    &  24,85 h          &  7,82 sec          &  20,76 h          &  6,53 sec       \\ \hline
10         & 8,008     &  11,69 h      	   &  5,25 sec       	&  3,54 h        	&  1,59 sec       \\ \hline
11         & 4,368     &  3,67 h       	   &  3,03 sec       	&  2,27 h        	&  1,89 sec       \\ \hline
12         & 1,820     &  29,81 min        &  0,98 sec       	&  22,12 min        &  0,73 sec       \\ \hline
13         & 560       &  7,42 min     	   &  0,79 sec          &  4,81 min       	&  0,51 sec       \\ \hline
14         & 120       &  31,56 sec    	   &  0,26 sec          &  18,68 sec     	&  0,15 s         \\ \hline
15         & 16        &  0,99 sec     	   &  0,06 sec          &  2,24 sec      	&  0,14 sec       \\ \hline
16         & 1         &  0,01 sec     	   &  0,01 sec       	&  0,01 sec      	&  0,01 sec       \\ \hline
\textbf{TOTAL}	&	\textbf{65,536}	&	\textbf{100,26 h}	&	\textbf{5,50 sec}	&	\textbf{64,49 h}	&	\textbf{3,54 sec}	\\ \hline
\end{tabular}
\end{table}

\begin{table}[ht]
\caption{Runtime comparison between $MPC$ and \textit{exact\_mig} by Depth}
\center
\label{CompRunTimeLvl}
{\renewcommand{\arraystretch}{1.2}
\begin{tabular}{|c|c|c|c|c|c|}
\hline 
\multicolumn{2}{|c|}{} & \multicolumn{2}{c|}{\textbf{MPC}}    & \multicolumn{2}{c|}{\textbf{exact\_mig}} \\ \hline
\textbf{Depth} & Total Functions & Total Runtime & Avg. Runtime & Total Runtime & Avg. Runtime    \\ \hline
\textbf{0} 	   & 10 			 & 0,12 sec		 & 0,01 sec 	& 0,23 sec		& 0,02 sec		   \\ \hline
\textbf{1} 	   & 80 			 & 1,07 sec		 & 0,01 sec		& 1,62 sec		& 0,02 sec		   \\ \hline
\textbf{2} 	   & 10,260 		 & 103,81 sec 	 & 0,01 sec 	& 318,34 sec	& 0,03 sec		   \\ \hline
\textbf{3} 	   & 55,184 		 & 91,80 h		 & 5,98 sec 	& 50,26 h		& 3,27 sec		   \\ \hline
\textbf{4}     & 2               & 8,46 h		 & 4,23 h 		& 14,22 h		& 7,11 h		    \\ \hline
\end{tabular}
}
\end{table} 

Note that even trought the $MPC$ can generate faster results for functions with 0, 1, 2 or 4 levels, in most cases it is still slower than \textit{exact\_mig}.

For $n=5$ a sample of 1,000 randomly generated functions were used and the $MPC$ algorithm was able to achieve lower cost results for $477 (48\%)$ functions, and equal cost results for $112 (11\%)$. 

The $MPC$'s total runtime for the generated sample was $11,62$ hours, with a average runtime of $41,63$ seconds. The \textit{exact\_mig}'s total runtime was $19,33$ hours, with a average runtime of $1,15$ minute. Therefore, the $MPC$ was able to generate results 66\% faster than \textit{exact\_mig}.  

Note that results for $n=3$ are not presented because both algorithms return optimal solutions for all $256$ possible functions.

\section{Conclusions} \label{conclusion}

In this paper we present the $MPC$ algorithm, which aims to generate majority functions based on a input truth table. We also present a study on the main concepts of majority boolean algebra and primitive functions. 
With the proposed cost criteria and for functions where $n=3$ or $n=4$, the $MPC$ presented, in the most part, results better or equal to \textit{exact\_mig}. 

For functions with $n=4$, from a total of 65,536 possible functions, the $MPC$ generated functions with lower cost in 42,987 (66\%) cases and functions with equal cost in 7,198 (11\%) cases.
For functions with $n=5$, from a sample of 1,000 functions, the $MPC$ found better results for 477 (48\%) functions and equal results for 112 (11\%). The $MPC$'s code is available on: https://github.com/EvandroFerraz/mpc. The list of functions used to compare $MPC$ and \textit{exact\_mig} for $5$-input functions can also be find in the link.

\bibliographystyle{splncs04}
\bibliography{mybibliography}

\end{document}